\documentclass[3p,12pt,number]{elsarticle}
\usepackage{fixltx2e}
\usepackage{latexsym}
\usepackage{array}
\usepackage{mathptmx}
\usepackage{bm}
\usepackage[htt]{hyphenat}
\usepackage{floatflt}
\usepackage[hyphens]{url}
\usepackage{color}
\usepackage{listings}
\usepackage{multirow}
\usepackage{booktabs}
\usepackage[lined,linesnumbered,nofillcomment]{algorithm2e}
\usepackage{listings}

\definecolor{gray}{rgb}{0.95,0.95,0.95}
 
\usepackage{graphicx}
\usepackage[cmex10]{amsmath}
\usepackage{mdwmath}
\usepackage{mdwtab}
\usepackage{eqparbox}
\usepackage[font=footnotesize]{caption,subcaption}
\usepackage{fixltx2e}
\newcommand{\GFLOPS}{GFLOPS}
\newcommand{\MB}{MB}

\newcommand{\GBS}{GB/s}

\newcommand{\MFLUP}{{\mbox{MFLUPS}}}
\newcommand{\MFLUPS}{{\mbox{\MFLUP{}}}}

\newcommand{\MLUPS}{\MFLUPS}

\newcommand{\NVIDIA}{{nVIDIA}}
\newcommand{\GTTWO}{GT--200}

\newcommand{\TESLA}{TESLA}

\newcommand{\JUROPA}{JUROPA}
\newcommand{\JUGENE}{JUGENE}
\newcommand{\NEC}{NEC Nehalem cluster}
\newcommand{\TINYGPU}{TinyGPU}

\newcommand{\PCIe}{PCIe}

\newcommand{\GPU}{GPU}
\newcommand{\GPUS}{\GPU{}s}
\newcommand{\GPUs}{\GPU{}s}
\newcommand{\CPU}{{CPU}}
\newcommand{\CPUS}{{CPUs}}
\newcommand{\CPUs}{{CPUs}}
\newcommand{\CUDA}{{CUDA}}
\newcommand{\bi}{\begin{itemize}}
\newcommand{\ei}{\end{itemize}}

\journal{Parallel Computing}

\begin{document}
\begin{frontmatter}


\title{A Flexible Patch-Based Lattice Boltzmann\\Parallelization Approach for Heterogeneous GPU--CPU Clusters}

\author[LSS]{Christian Feichtinger\corref{cor1}}
\ead{Christian.Feichtinger@informatik.uni-erlangen.de}
\author[RRZE]{Johannes Habich}
\author[LSS]{Harald K\"ostler}
\author[RRZE]{Georg Hager}
\author[LSS]{Ulrich R\"ude}
\author[RRZE]{Gerhard Wellein}

\cortext[cor1]{Corresponding author}

\address[LSS]{Chair for System Simulation\\
   University of Erlangen-Nuremberg}
\address[RRZE]{Regional Computing Center Erlangen\\
   University of Erlangen-Nuremberg}
\begin{abstract}
Sustaining a large fraction of single GPU performance in parallel
computations is considered to be the major problem of GPU-based clusters.
In this article, this topic is addressed in the context of a lattice
Boltzmann flow solver that is integrated in the WaLBerla software framework.
We propose a multi-GPU implementation using a
block-structured MPI parallelization, suitable for load balancing
and heterogeneous computations on CPUs and GPUs. 
The overhead required for multi-GPU simulations is discussed in detail
and it is demonstrated that the kernel performance can be sustained to a large extent.
With our GPU implementation,
we achieve nearly perfect weak scalability on InfiniBand clusters.
However, in strong scaling scenarios multi-GPUs make less efficient use of the hardware than
IBM BG/P and x86 clusters. Hence, a cost analysis must determine the best course of action
for a particular simulation task.
Additionally, weak scaling results of heterogeneous simulations 
conducted on CPUs and GPUs simultaneously are presented using clusters equipped
with varying node configurations.
\end{abstract}


\end{frontmatter}
%
%
%
%
%

\section{Introduction}

In the field of computational fluid dynamics (CFD), flow solvers based
on the lattice Boltzmann method (LBM) have become a well-established
alternative for solving the Navier-Stokes equations directly. The LBM algorithm is a
cellular automaton derived from the Boltzmann equation; each node
({\em cell}) on the computational grid exchanges information with its
neighbors, which makes memory bandwidth the performance-limiting
bottleneck of the LBM in most cases. 
Modeling real systems requires large computational
effort, therefore performance optimization and parallelization of LBM
codes are very active fields of research. \GPU\ architectures offer the
highest memory to processor chip bandwidth available today in commodity hardware and promise
a big performance gain for memory-bound applications.
However, tremendous effort has to be put into highly efficient
LBM codes even on single \GPU{}s 
\cite{2008:Toelke:Accuracy,gpu:Habich:2009:paper,gpu:obrecht2010}.
First promising results of nonregular implementations
\cite{gpu:Bernaschi:2010} show that the LBM is applicable to 
nonuniform domains and multi-\GPU\ clusters as well.

In order to push these experimental efforts into real production
CFD applications it is crucial to establish scalable LBM codes on \GPU\
clusters. Still the current Top500 list \cite{hpc:TOP500List}
contains only a few \GPU\ clusters, since the nonstandard programming
paradigm and the rather slow \CPU-to-\GPU\ connection are 
obstacles that hamper their general applicability.
The main contribution of this paper is to show that it is possible to
exploit the full computational power of currently emerging \GPU--\CPU\
clusters.  We start by applying low-level optimizations to the \GPU\
kernels to improve single-\GPU\ performance, then move to multiple
\GPU{}s using MPI-based distributed memory parallelization, and
finally establish load-balanced heterogeneous \GPU--\CPU\ parallelism
by incorporating the otherwise idle multicore CPUs and \GPU-less
compute nodes into the flow solver.

To bring together high performance and high productivity we apply a
{\em Patch and Block} design, which divides the computational domain
into subregions that are distributed to the {\em compute units} (\GPU{}s
or teams of \CPU\ threads in a multicore node). This is a choice as
to how many subregions are assigned to each compute unit instead of individual cells, 
simplifying static load balancing. Ghost layers are exchanged between
neighboring subregions using the appropriate data paths (shared memory,
PCIe bus, InfiniBand interconnect). A simulation is thus able
to run in parallel on a heterogeneous cluster comprising various
different architectures. The whole code is included in the
WaLBerla (Widely applicable lattice Boltzmann solver from Erlangen) software framework, which is employed in many CFD
applications. We can show that high computational performance can be
sustained within WaLBerla and therefore only very small application
management and communication overhead is added.

This paper is organized as follows. We start with a brief description
of the architectures used for measurements and the LBM method in
Sec.~2.  Code optimizations and performance models are shown in
Sec.~3. Section~4 describes the WaLBerla software framework and the
heterogeneous parallelization approach.  In Sec.~5 we finally
analyze the performance behavior of our code on single \CPU{}s, single
\GPU{}s, and heterogeneous \CPU--\GPU\ clusters in strong and weak
scaling scenarios.

%
%
%
%
%
%
\section{Methods and Architectures}
\subsection{The Lattice Boltzmann Method}
%
%
%
%
%
%
%
The LBM has evolved over the last two decades and is today widely
accepted in academia and industry for solving incompressible flows.
Coming from a simplified gas-kinetic description, i.e. a velocity
discrete Boltzmann equation with an appropriate collision term, it
satisfies the Navier-Stokes equations in the macroscopic limit with
second order accuracy~\cite{lbm:chen:1998,lbm:succi:2001}. In contrast
to conventional computational fluid dynamic methods, the LBM uses a
set of particle distribution functions (PDF) in each cell to describe
the fluid flow.  A PDF is defined as the expected value of particles
in a volume located at the lattice position $\vec {x}$ with the
lattice velocity $\vec{e}_{i}$.
Computationally, the LBM is based on a uniform grid of cubic cells
that are updated in each time step using an information exchange with
nearest neighbor cells only. Structurally, this is equivalent to an
explicit time stepping for a finite difference scheme.  For the LBM
the lattice velocities $\vec{e}_{i}$ determine the finite difference
stencil, where $i$ represents an entry in the stencil.  Here, we use
the so-called D3Q19 model resulting in a $19$ point stencil 
and 19 PDFs in each cell.  The evolution of a single PDF $f_i$ is
described by
\begin{eqnarray}
  \label{eq:f_coll}
\nonumber
&f_i(\vec{x}+\vec{e}_i\Delta t, t+\Delta t) =
  f^{\mathrm{coll}}_i(\vec{x}, t) = &\\[2mm]
\label{eq:lbm}
&\ \ \ \ \ \ f_i(\vec{x}, t) - \frac{1}{\tau} \big[
   f_i(\vec{x},t)
- f_i^{\mathit{eq}}\left(\rho(\vec{x}, t), \vec{u}(\vec{x}, t)\right) \big]\qquad&\\[2mm]
\label{eq:feq}
&f_i^{\mathit{eq}}\left(\rho(\vec{x}, t), \vec{u}(\vec{x}, t)\right)=
w_i
\left[ 
\rho + \rho_0 
\left(
3\vec{e}_i\vec{u} + 4.5 (\vec{e}_i\vec{u})^2 -1.5 \vec{u}^2
\right) 
\right] 
&\\[2mm]
 & i=0\ldots 19,& \nonumber
\end{eqnarray}
and can be split into two steps: A collision step applying the
collision operator and a propagation step advecting the PDFs to the
neighboring cells. In this article, we use the single relaxation time
collision operator~\cite{lbm:succi:2001}.  In Eq. \ref{eq:lbm},
$f_i^{\mathrm{coll}}$ denotes the intermediate state after
collision but before propagation. The relaxation time $\tau$ can be
determined from the kinematic viscosity $\nu=( \tau
-\frac{1}{2})c_s^2\delta t$, with $c_s$ as the speed of sound.
Further, $f^{eq}_i$ is a Taylor expanded version of the
Maxwell-Boltzmann equilibrium distribution
function~\cite{lbm:succi:2001} optimized for incompressible
flows~\cite{lbm:He:1997_2}.  For the isothermal case, $f^{eq}_i$
depends on the macroscopic velocity~$\vec{u}(\vec{x}, t)$ and the
macroscopic density~$\rho(\vec{x}, t)$, and the lattice weights $w_i$
are $\frac{1}{3}$, $\frac{1}{18}$ or $\frac{1}{36}$.  The macroscopic
quantities $\rho$ and $\vec u$ are determined from the $0^{th}$ and
$1^{st}$ order moment of the distribution functions $ \rho(\vec{x},
t)= \rho_0 + \delta \rho(\vec{x}, t) = \sum_{i=0}^{18}f_i(\vec{x}, t),\nonumber $
and $\rho_0 \vec{u}(\vec{x}, t)=\sum_{i=0}^{18} {\vec e}_{i}
f_i(\vec{x}, t)$, where $\rho$ is split into a constant part $\rho_0$ and a
slightly changing perturbation $\delta \rho$.  The equation of state
of an ideal gas provides the pressure $p(\vec{x},t) = c_s^2 \rho
(\vec{x},t)$.
\par
Usually, the PDFs are initialized to $f_i^{\mathit{eq}}\left(\rho_0,
  0\right)$.  To increase the accuracy of simulations in single
precision we use $\tilde{f}_i(\vec{x}, t)=f_i(\vec{x},
t)-f_i^{\mathit{eq}}\left(\rho_0, 0\right)$ and
$\tilde{f}_i^{\mathit{eq}}\left(\rho(\vec{x}, t), \vec{u}(\vec{x},
  t)\right) = f_i^{\mathit{eq}}\left(\rho(\vec{x}, t),
  \vec{u}(\vec{x}, t)\right) - f_i^{\mathit{eq}}\left(\rho_0, 0\right)
$
as proposed by~\cite{lbm:He:1997_2}
resulting in PDF values centered around
$0$. According to~\cite{2008:Toelke:Accuracy}, it is possible with the LBM
scheme described above to achieve accurate single precision results,
which is important for GPU implementations.
\par
A further important issue is the implementation of the propagation,
for which there exist two schemes: First a pushing and second a
pulling.  For the first case, the PDFs in a cell are first collided
and then pushed to the neighborhood. In the second case, the
neighboring PDFs are first pulled into the lattice cell and then
collided. Additionally, the
propagation step introduces data dependencies to the LBM, which
commonly result in an implementation of the LBM using two PDF grids.
However, these dependencies are of local type, as only PDFs of
neighboring cells are accessed. Hence, the LBM is particularly well
suited for massively parallel simulations~
\cite{hpc:Zeiser:2009:ppl:paper,el00290,Feichtinger:2009:ParComp}.  In
WaLBerla, we use the {\em pull} approach as it is better suited
for our parallelization (see Sec. \ref{sec:walberla} for details
on the parallelization).
\par
The most common approach for implementing solid wall boundaries in the LBM is the bounce-back
(BB) rule~\cite{lbm:chen:1998,lbm:succi:2001}, i.e. if a distribution
is about to be propagated into a solid cell, the distribution's
direction is reversed into the original cell.  BB generally assumes
that the wall is in the middle between the two cell centers, i.e. half-way.
This formulation leads to: $ f_{\bar{i}} (\vec{x},t +
\Delta t) = f_i(\vec{x},t) + 6w_i\rho_0\vec{e}_i\vec{u}_w $ with
$\vec{e}_{\bar{i}} = -\vec{e}_i$ and $u_w$ being the velocity
prescribed at the wall.
\subsection{Hardware Environments}
\label{sec:hardware}

\subsubsection{\CPU{}-based Cluster Systems}
In general, current clusters based on dual-socket Intel quad-core
processors offer a peak node performance in the range of $60$ to $100$
\GFLOPS{}. The on-chip memory controllers with up to three DDR3 memory
channels per socket provide a theoretical peak node bandwidth of
$64$~\GBS{}.
%
%
%
The clusters introduced in the following table all share this common architecture:\\[5mm]
{
\footnotesize
\centering
\begin{tabular}{llclccc}\toprule[0.3mm]
   &Nodes&Processor&Interconnect&Clock Speed&Memory&\NVIDIA{} GPUs per Node\\
   & & Xeon   &         &[GHz]      &[GB]&\\
\midrule[0.3mm]
\TINYGPU{}~\cite{sys:tinygpu}&8&X5550&DDR IB&2.66&24&2 x \TESLA{} C1060\\
\JUROPA{}~\cite{sys:juropa}&2208&X5570&QDR IB&2.93&24&\\
NEC Nehalem~\cite{sys:nec-hlrs}&700&X5560&DDR IB~$^*$&2.8&12&2 x \TESLA{} S1070 \\
   \multicolumn{3}{l}{ \tiny $^*$~Oversubscribed IB backbone}&&  &&(30 nodes)\\
\bottomrule[0.3mm]
\end{tabular}
}\\[5mm]
The IBM BlueGene/P-based cluster \JUGENE{}~\cite{sys:jugene} comprises $73728$ compute
nodes, each equipped with one $850$~MHz PowerPC $450$ quad-core processor and
$4$ GB memory, which are connected via a proprietary high speed
interconnect offering $850$~MB/s per link direction.
\subsubsection{\NVIDIA{} Graphic Processing Units} 
The \GTTWO{}-based \GPUS{} are the second generation of \NVIDIA{}
graphics cards capable of GPGPU computing using the
{\em Compute Unified Device Architecture} (\CUDA{})~\cite{gpu:nvidiaCudaTK:2009}. A \GPU{} has several multiprocessors
(MP), each with $8$ processor cores. Computations are executed
by so-called threads, whereas up to $1024$ threads are concurrently
running on one MP in order to hide memory latency by efficient
scheduling. Threads are organized in \GPU{}-blocks, which are pinned
to an MP over the whole runtime. Each MP has $16384$ registers and
$16$~kB of shared memory available, i.e. there are only $16$ registers
and about $16$~bytes per thread if $1024$ threads are running in
parallel. Hence, the concurrency is limited if kernels allocate more than $16$ registers, which 
has a severe impact on performance.
See~\cite{gpu:nvidiaCudaProgrammingGuide2.3.1:2009} for
further details on \CUDA{} and \NVIDIA{} \GPU{} hardware. 
\subsection{Interconnects}
\begin{figure}[htb]
\centering
\includegraphics[clip,width=0.5\columnwidth]{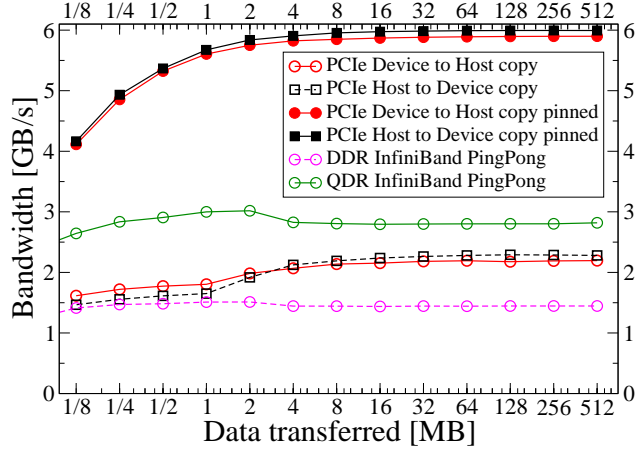}
\caption{Host--\GPU{} and MPI PingPong bandwidth measurements on
  \TINYGPU{}. The function {\em cudaMemcpy} implementing a vector copy
  is used for all \PCIe{} copy operations.}
\label{plot:PCIEBW}
\end{figure}
Most of today's high performance systems use InfiniBand (IB) for the
connection of the compute nodes. Heterogeneous computations on \CPUS{} and
\GPUS{} require PCI-Express (\PCIe{}) transfers for both IB and
\CPU{}--\GPU{} communication. In order to develop a sensible
performance model, all involved communication paths must be
considered.
\par
\PCIe{} 2.0 x16 is currently the fastest peripheral bus with
a peak transfer bandwidth of 8~\GBS{} per direction. 
Figure~\ref{plot:PCIEBW} shows that one can maintain about 6~\GBS{}
bandwidth if the transfered data is larger than $2$~\MB{} and the \CUDA{} 
call {\em cudahostalloc} is used to allocate so-called {\em pinned memory} on
the host. Pinned memory in contrast to memory allocated by {\em
  malloc} will not be paged out, is private to the process allocating
it, and is local to the physical socket of the allocating process. The
advantage of pinned memory results from the possibility to use fast
direct-memory-accesses (DMA).  With our current LBM implementation
packets in the range of $250$~kB to $500$~kB are exchanged per \PCIe{}
data transfer, leading to an effective bandwidth between about $5$~\GBS{}
and $6$~\GBS{}. 
Please note that two GPUs on the \NEC{} have to share the same \PCIe{} bus, which is
capable of transferring $12.8$~\GBS{}.
\par
IB host adapters are connected to the host via
the \PCIe{} x8 interface. IB bandwidth measurements of the
Intel IMB {\em PingPong} benchmark~\cite{hpc:intel:mpibench} for quad-(QDR)
and double-(DDR) data rate IB can be found in
Fig.~\ref{plot:PCIEBW}. The measurements show that QDR ($3.0$~GB/s) doubles the
bandwidth compared to DDR ($1.5$~GB/s) and that the \GPU{}'s \PCIe{} operates with at
least twice the bandwidth.
\par
The performance of LBM codes is usually given in terms of {\em million fluid
lattice cell updates per second} (MFLUPS) instead of GFlops, as
the actual executed GFlops cannot be determined precisely. 
Table~\ref{tab:HybridComputingEstimate} gives an estimate for the
minimal impact of the data transfer over all interconnects on
performance. The compute time of the kernel $t_k$ and the IB and \PCIe{} data transfer times $t_t$
can hereby be determined by 
\[
 t_{k} =   \frac{n_{cell}^3}{P}\ \ \ \ \mbox{and} \ \ \ \
 t_{t} = \frac{ 2 \cdot n_{cell}^2 \cdot n_{PDF} \cdot n_{plane} \cdot s_{PDF}}{B},
\]
where $P$ is the performance, $n_{Cell}$ the number of lattice cells per dimension, $n_{PDF}$ 
the number of PDFs communicated per boundary cell,
$n_{plane}$ the number of planes to be communicated, $s_{PDF}$ the size in bytes of a PDF
and $B$ is the bandwidth of 
the corresponding interconnect.
It was assumed that all domain boundaries have to be communicated,
which results in the transfer of $6$ boundary planes with $5$ PDFs per cell.
\begin{table}[htb]
\footnotesize
\centering
\begin{tabular}{llrl}
  \toprule[0.3mm]
  Steps &\multicolumn{3}{c}{ Tesla C1060 ($\sim 300$ \MFLUPS{})} \\
  \midrule
  Compute Time && 3.3 ms&\\
  PCIe: 5 \GBS{} &(I)& 0.48 ms& \\
  IB: 3.0 \GBS{} &(II)& 0.8 ms&\\
  \midrule
  Total Time: & (I)  &\multicolumn{2}{r}{ 3.78 ms $\rightarrow$ 264 \MLUPS{}}\\
  &(I+II) &\multicolumn{2}{r}{ 4.58 ms $\rightarrow$ 218 \MLUPS{}}\\
  \bottomrule[0.3mm]\\
\end{tabular}
\caption{\label{tab:HybridComputingEstimate} 
  Performance estimates for multi-GPU single precision LBM simulations
  including InfiniBand and \PCIe{} transfers of the complete boundary data.
  The estimated times are based on the obtainable bandwidth.
  A domain size of $100^3$ lattice cells is assumed for the example and the pure kernel 
  performance has been taken from Fig.~\ref{plot:MultiBlockSingleGPU-Kernel}.}
\end{table}
%

%
%
%
%
%
%
%
\section{\CPU{} and \GPU{} Kernel Implementation}
\label{sec:kernels}
\subsection{Upper Bound Performance Estimation}
The performance of our LBM implementation is like most scientific
codes dominated by memory bandwidth. To estimate an upper bound for
the obtainable LBM bandwidth on CPUs, we employ the vector operation
$c(:)=a(:)$ from the {\em STREAM
Benchmarks}~\cite{hpc:streambench:2010}, which results in a memory
bandwidth of $33$ GB/s on \JUROPA{}. The \CPU{}'s cache hierarchy and
arithmetic units are fast enough so that computations and in-cache transfers
are completely hidden by memory loads and stores.  For the GPU
bandwidth, we implemented our own benchmark, which achieved a maximum
memory bandwidth of $78$ GB/s on a \NVIDIA{} \TESLA{} C1060, if the
occupancy is at least $0.5$, i.e. if at least $512$ threads out of the
maximum of $1024$ (\GTTWO{}) threads are scheduled per MP.  Further
benchmark details can be found in~\cite{gpu:Habich:2009:paper}.
Furthermore, the bytes transferred for each LBM lattice cell
update $n_{bytes}$ can be determined by~\cite{2006:Wellein:CompFluids}
\[n_{bytes} = n_{stencil} \cdot (n_{loads}+n_{store})
\cdot s_{PDF},\] where $n_{stencil}$ is the size of the LBM stencil, and $n_{loads}$ and $n_{stores}$ the number of load and stores.
Due to the {\em Read-for-Ownership},
this results in $228$ bytes using single precision (SP) and $456$
bytes using double precision (DP) for the CPU, and $152$ (SP) / $304$
(DP) bytes for the GPU implementation. Thus, it is possible 
to estimate an upper limit for the LBM node performance. For one node on \JUROPA{} we
estimate a performance of $144$ (SP) / $72$ (DP) \MFLUPS{} and $516$
(SP) / $258$ (DP) \MFLUPS{} for one \NVIDIA{} \TESLA{} C1060.
\subsection{Kernel Performance and Implementation Details}
\begin{figure}[!t]
\centering
\includegraphics[clip,height=6cm]{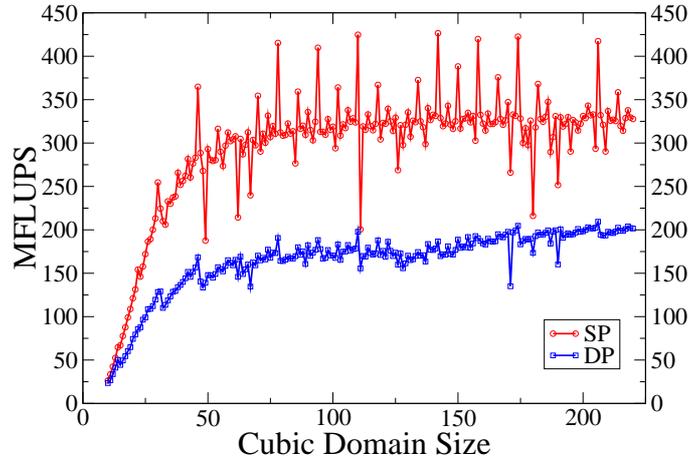}
\caption{Single-\GPU{} measurements of the pure \GPU{} kernel
           performance for SP and DP on a \NVIDIA{} \TESLA{} C1060.
         }
\label{plot:MultiBlockSingleGPU-Kernel}
\end{figure}  
One key aspect for achieving a good LBM kernel performance is the data
layout.  There exist two major implementation strategies: The
Array-of-Structure (AoS) and the Structure-of-Arrays (SoA) layout.
For the AoS layout, the PDFs of each cell are stored adjacent in
memory, whereas for the SoA Layout the PDFs pointing in the same
lattice direction are adjacent in memory. Our CPU kernel
implementation uses the AoS layout together with the pull streaming
approach, and to improve the performance, arithmetic optimizations have been applied.  
In addition, the Patch and Block data structures
introduced in Sec.~\ref{sec:patches} allow for the decomposition of
the simulation domain into smaller subdomains, leading to an implicit
spatial blocking.
No further unrolling or
spatial and temporal blocking is applied. Our implementation reaches
up to $78$~(SP)~/~$55$~(DP) MFLUPS on one node of \JUROPA{} and up to
$7.3$~(SP)~/~$6.1$~(DP) on one node of \JUGENE{}. This is slightly
lower, but comparable to well-optimized solvers,
e.g.~\cite{hpc:Zeiser:2009:ppl:paper}.  The DP kernel is about 23~\%
off from the performance estimated before and still in agreement with
the model. The large discrepancy of nearly 50~\% for the SP kernel can be
attributed to the computational intensity of the nonvectorized LBM
kernel, making the code essentially not memory, but computationally
bound.
\par
In contrast to the CPU implementation, the \GPU{} implementation uses
the SoA layout, because in combination with the
pull streaming approach it is possible to align the memory
writes. In addition, the scattered loads that occur in our
implementation can be efficiently coalesced by the memory subsystem.
Hence, we do not have to use the shared memory of the GPU.
For the scheduling of the threads, we adopted a scheme first proposed
in~\cite{2008:Toelke:Accuracy}, where each GPU thread updates one
lattice cell and one GPU block is assigned one row of the simulation
domain.
In order to improve the kernel performance, we reduced the number of
registers used for each thread by prefetching the PDFs into temporal
variables and also by modifying the array accesses as described
in~\cite{gpu:Habich:2009:paper}.  With these optimizations, we can
achieve a maximum occupancy of $0.5$. The maximum performance for some
domain sizes has been around $500$~(SP)~/~$250$~(DP), which agrees well
with our performance estimates and also with the results
in~\cite{gpu:obrecht2010}. A comparison to~\cite{2008:Toelke:Accuracy}
is rather difficult as they used a different LBM stencil and
hardware has evolved.  Still, the sustained memory bandwidth of both
implementations on the particular hardware is around $70$~\% of peak
bandwidth.  A detailed kernel performance analysis for cubic domain
sizes is depicted in Fig.~\ref{plot:MultiBlockSingleGPU-Kernel}.  The
measured performance fluctuations for varying domain sizes result from
the different numbers of scheduled threads per MP and from memory
alignment issues.
%

%
%
%
%
%
\section{The WaLBerla Framework}
\label{sec:walberla}
%
%
WaLBerla is a massively parallel multiphysics software framework 
that is originally centered around the LBM, but
whose applicability is not limited to this algorithm.
Its main design goals are to provide excellent application performance
across a wide range of computing platforms and the easy integration of
new functionality.  In this context additional functionality can
either extend the framework for new simulation tasks, or optimize
existing algorithms by adding special-purpose hardware-dependent
kernels or new concepts such as load balancing strategies.
In order to achieve this flexibility, WaLBerla has been designed
utilizing software engineering concepts such as the spiral model and
prototyping~\cite{boehm1986spiral,van2000software}, and also using
common design patterns~\cite{gamma1995design}.
Several researchers and cooperation partners have already used the
software framework to solve various complex simulation tasks. Amongst
others, free-surface flows~\cite{2009:Donath:EuroPar} using a
localized parallel algorithm for bubbles coalescence, free-surface
flows with floating objects~\cite{ma:bogner}, flows through porous
media, clotting processes in blood vessels~\cite{ma:haspel},
particulate flows for several million volumetric
particles~\cite{2010:Goetz:ParCo} on up to $8192$ cores, and a
fluctuating lattice Boltzmann~\cite{PhysRevE.76.036704} for nano
fluids have been included.  In addition to the strictly Eulerian view
of field equations and their discretization, WaLBerla also supports
Lagrangian representations of physical phenomena, such as
e.g. particulate flows.
Currently, the prototype WaLBerla~$2.0$ is under development extending
the framework for heterogeneous simulations on \CPUs{} and \GPUs{}, and load
balancing strategies.  Heterogeneous computations are already
supported, but the designs for dynamic load balancing strategies are
currently under development, although the underlying data structures
can already be used for static
load balancing.
\par
In WaLBerla, all simulation tasks are broken down into
several basic steps, so-called {\em Sweeps}.  A Sweep can be divided
into two parts: a communication step fulfilling the boundary
conditions for parallel simulations by nearest neighbor communication
and a communication independent work step traversing the process-local
grid and performing operations on all cells.
The work step usually consists of a kernel call, which is realized for instance by a function object
or a function pointer.  As for each work step there may exist a list of
possible (hardware dependent) kernels, the executed kernel is selected
by our functionality management (see below).
For pure LBM simulations only one Sweep is needed exchanging PDF
boundary data during the communication phase and executing one of the kernels 
that have been described in Sec.~\ref{sec:kernels}.
The functionality management in WaLBerla $2.0$ selects the required
kernels according to meta data provided with each kernel.  This data allows
the selection of different kernels for different simulation runs,
processes and subregions of the simulation domain, so-called {\em Blocks}
(see Sec.~\ref{sec:patches}). Hence, it is possible to
specifically select, for heterogeneous computations even on each
single process, hardware optimized kernels.
Further details on the functionality management can be found in Sec.~\ref{sec:func_concept}.
\par
A further fundamental design of the whole software framework is our
{\em Patch} and {\em Block} data structure, which is a specific
version of block-structured grids. Besides forming the basis for the
parallelization and load balancing strategies, Blocks are also
essential to configure the domain subregions with regard to the
simulated task and the utilized hardware.  More information on the
Patch and Block data structure can be found in Sec.~\ref{sec:patches}.
Further, WaLBerla enables parallel MPI simulations of various
simulation tasks. In order to do so, the process-to-process
communication supports messages, containing data from any kind of data
structure conforming to a documented interface, of arbitrary length
and data type as well as the serialization of messages to the same
process. Using our parallelization it is possible to represent even
complex communication patterns, such as our localized bubble merge
algorithm~\cite{2009:Donath:EuroPar} or our parallel multigrid solver
ported from~\cite{Koestler08}.
The general parallelization design is described in
Sec.~\ref{sec:paralleization}.  For parallel simulations on \GPUs{},
the boundary data of the \GPU{} has first to be copied by a \PCIe{}
transfer to the \CPU{} and then be communicated via the MPI
parallelization. Therefore, the data structures of the single core
implementation are extended by buffers on \GPU{} and \CPU{} in order
to achieve fast \PCIe{} transfers. In addition, on-\GPU{} copy kernels
are added to fill these buffers.  In Sec.~\ref{sec:multi-gpu} the
details of our parallel \GPU{} implementation are introduced. To
support heterogeneous simulations on \GPUs{} and \CPUs{}, we execute
different kernels on \CPU{} and \GPU{} and also define a common
interface for the communication buffers, so that an abstraction from
the hardware is possible.  Additionally, the work load of the CPU and
the GPU processes has to be balanced. In our approach this is achieved
by allocating several Blocks on each GPU and only one on each CPU-only
process.
\subsection{Functionality Management}
\label{sec:func_concept}
The functionality management in WaLBerla $2.0$ allows to select
different functionality (e.g. kernels, communication functions) for
different granularities, e.g. for the whole simulation, for
individual processes, and for individual Blocks. This is realized by
adding meta data to each functionality
consisting of three unique identifiers (UID).
\par
\begin{center}
\footnotesize
\begin{tabular}{llll}\toprule[0.3mm]
UID & Name &  Granularity& Example\\
\midrule
{\em fs} & Functionality Selector & Simulation & Gravity on/off\\
{\em hs} & Hardware Selector & Process& CPU and/or \GPU{}\\
{\em bs} & Block Selector & Block & LBM\\
\bottomrule[0.3mm]
\end{tabular}\\[6mm]
\end{center}
On the basis of these UIDs the kernels can be selected according to
the requirements of the simulated scenarios. Hence, physical effects
can be turned on/off in an efficient well-defined manner by means of
the {\em fs} selector.  Hardware-dependent kernels can be selected for
different architectures depending on the {\em hs} selector and
simulation tasks can be selected via the {\em bs} selector.  A complex
example for the capabilities of our concept are heterogeneous LBM
simulations on \CPU{}s and GPUs described in
Sec.~\ref{sec:hetero_impl}.
\subsection{Patch and  Block Concept}
\label{sec:patches}
\begin{figure}[htbp]
\centering
\includegraphics[width=0.5\columnwidth]{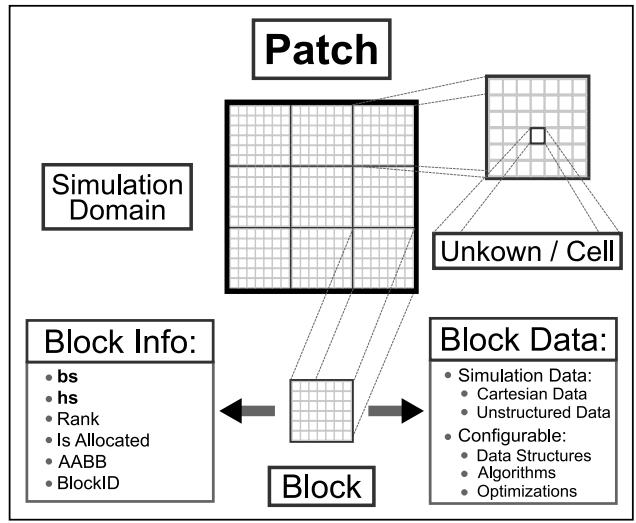}
\caption{Patch and Block Design. Each Block stores management
  information consisting of a block and a hardware selector, a MPI
  rank, an axis aligned bounding box (AABB), and a Block identifier
  (BlockID) required for the identification of individual Blocks.  The
  Blocks' management information is stored on all processes, but
  simulation data is only allocated on processes which are responsible
  for that particular Block.  }
\label{fig::patch_concept}
\end{figure}
In WaLBerla the simulation domain is described with our Patch and
Block design, which is illustrated in
Fig.~\ref{fig::patch_concept}. It has been developed in order to
support massively parallel simulations, load balancing strategies and
the configuration to simulation tasks and hardware.
A Patch hereby is a rectangular cuboid describing a region in the
simulation that is discretized with the same resolution.  In principal, these
Patches can be arranged hierarchically for grid refinement techniques,
but in this work we are using only one Patch covering the whole
simulation domain.
This Patch is further subdivided into a Cartesian grid of Blocks,
again of cuboidal shape, containing the actual grid-based data for the
simulation (simulation data).  With the aid of these Blocks the
simulation domain can be partitioned for parallel simulation.  It is
hereby possible to allocate several Blocks on a process in order to
support load balancing strategies. Additionally, with the help of the
functionality management the Blocks' data can be configured for the
simulated scenario.  In particular, each Block contains two kinds of
data: management information and simulation data.
The management data contains a {\em rank} parameter, which decides on
which process the simulation data of the Block is
allocated. Additionally, a hardware selector ({\em hs}) describes the
hardware on which the Block is allocated, whereby all Blocks on the
same process have the same hardware selector assigned to them.
Further, the management data contains a block selector ({\em bs})
deciding which task is simulated on a Block.  For the simulation data
each block stores a dynamic list of base class pointers. For
multiphysics simulations this allows to store an arbitrary number of
data fields, e.g. grid-based data for velocity, temperature or
potential values or unstructured particle data for particulate flows.
Hence, each block can be configured in the following way: During the
initialization of a simulation WaLBerla creates lists of possible
simulation tasks, kernels for each Sweep and several simulation data
types, whereby each entry in a list is connected to meta data for the
functionality management.  With the help of the selectors stored in the
management information it is possible to select 
which task has to be simulated, which simulation data has to be
allocated, and which kernels have to be selected for the Sweeps from these lists.
\subsection{General Design of the MPI Communication}
\label{sec:paralleization}
\begin{figure}[htbp]
\centering
\includegraphics[clip,width=0.6\columnwidth]{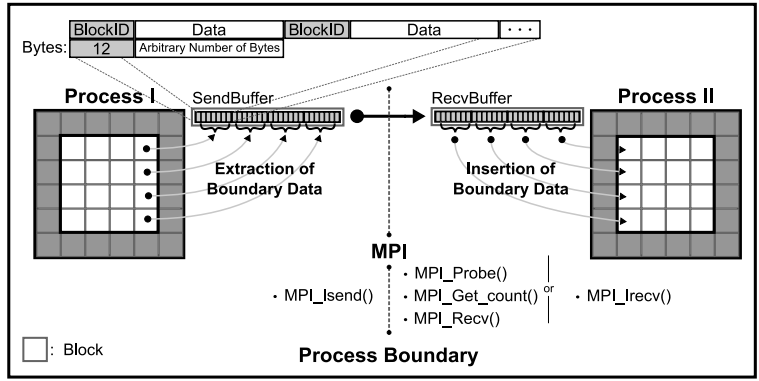}
\caption{Design for parallel simulations. In the Figure,
the MPI communication from process I to process II is depicted.
First, the data to be communicated is extracted with provided functions from each Block and stored in send buffers. 
For pure fluid flows only PDFs have to be sent.
On the sending side an MPI\_Isend is scheduled and on the receiving side the 
message is either received with a MPI\_Probe, MPI\_GetCount and a MPI\_Recv, or a MPI\_Irecv.
Note, that we attach a header to each Block message containing the
BlockID and a communication direction. This is required in order to
determine the Block to which the data has to be copied on
the receiving side.
}
\label{fig:paracomm}
\end{figure}
%

%
The parallelization of WaLBerla, which is depicted in Fig.~\ref{fig:paracomm}, 
can be broken down into three steps: a data extraction step, a MPI
communication step and a data insertion step.  During the data
extraction step, the data that has to be communicated is copied from
the simulation data structures of the corresponding Blocks. Therefore,
we distinguish between process-local and MPI communication for Blocks
lying on the same or different processes.
Local communication directly copies from the sending Block to the
receiving Block, whereas for the MPI communication the data has first
to be copied into buffers. For each process to which data has to be
sent, one buffer is allocated.
With the buffers, all messages from Blocks (block message) on the same process to another process
are serialized.  Additionally, the buffers are of data
type {\em byte} and thus the MPI messages can contain any data type
that can be converted into bytes.
To extract the data to be communicated from the simulation data,
extraction function objects are used.  For each communication step and
for each simulation data type several possible function objects are
provided during the configuration of the communication.  These are
again selected via the functionality management.
During the MPI communication one MPI message is sent to each process
waiting for data from the current process.  Therefore, nonblocking
MPI functions are used, if the message size can be determined a
priori. The data insertion step is similar to the data extraction,
only here we traverse the block messages in the communication
buffers
instead of the Blocks.
\begin{figure}[htbp]
\centering
   \includegraphics[width=0.8\columnwidth]{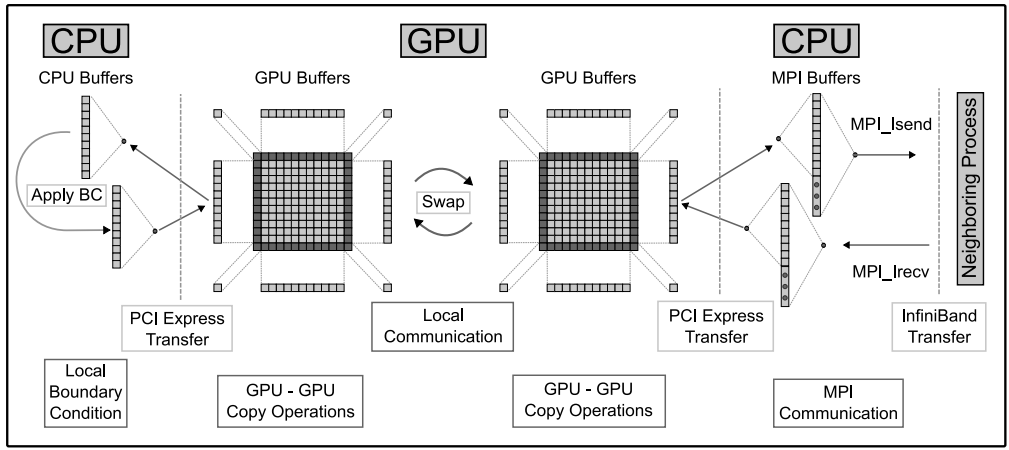}
\caption{Multi-GPU design.}
\label{fig:gpuImpl}
\end{figure}
\subsection{Multi \GPU{} Implementation}
\label{sec:multi-gpu}
For parallel \GPU{} simulations part of the data stored on the \GPU{}
has to be transferred to the \CPU{} via \PCIe{} transfers before it
can be communicated by means of the MPI communication. An efficient
implementation of this transfer is important in order to sustain a
large portion of the kernel performance. Hence, we only
transfer the minimum amount of data necessary, the boundary values of
the PDFs. Our parallel \GPU{} implementation is depicted in
Fig.~\ref{fig:gpuImpl} for one process having two Blocks.
It can be seen that we extended the data structures by additional
buffers on the \GPU{} and on the \CPU{} side. In $3$D, we add $6$ planes 
and $12$ edge buffers.  To update the ghost layer of the PDFs and to
prepare the \GPU{} buffers for the MPI communication additional
on-\GPU{} copy operations are needed.  The data of the buffers is
copied to the ghost layer of the Blocks before the kernel call and the
PDF boundary values of the PDF data are copied into the \GPU{} buffers
afterwards.
For parallel simulations, the MPI implementation of
Sec.~\ref{sec:paralleization} is used.  Here, the only difference to
the \CPU{} implementation are the extraction and insertion functions,
which for the local communication simply swap the \GPU{} buffers,
whereas the function {\em cudaMemcpy} is used to copy the data
directly from the \GPU{} buffers into the MPI buffers and vice versa
for the MPI communication.  To treat the boundary conditions at the
domain boundary, the corresponding GPU buffers are transferred via
{\em cudaMemcpy} to the CPU buffers. Next, the boundary conditions are
applied and the data is copied back into the GPU buffers. The boundary
conditions are fulfilled before the on-GPU copy operations.
\subsection{Heterogeneous \GPU{} / \CPU{} Implementation}
\label{sec:hetero_impl}
\begin{figure}[htbp]
\centering
\includegraphics[clip,width=0.6\columnwidth]{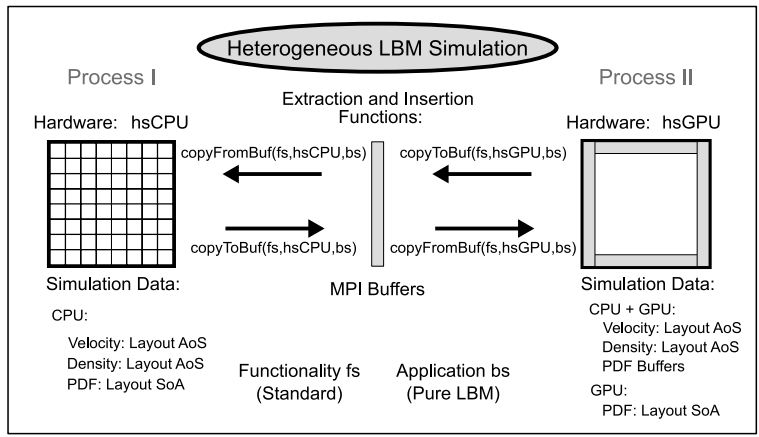}
\caption{Heterogeneous simulation on \GPU{} and \CPU{}.
The illustrated simulation is executed on two processes each having
one Block covering half of the simulation domain.
A standard LBM simulation is chosen as {\em fs} and on all Blocks the {\em bs} pure LBM is activated.
Further, the first process runs on a \CPU{} ({\em hsCPU}), whereas the second uses a \GPU{} ({\em hsGPU}).
According to these UIDs, the simulation data is allocated and the kernels, the extraction and insertion functions are selected.
For the communication, extraction functions {\em copyToBuf} are selected by the UIDs of the corresponding
Block to copy the communicated data in the specified format into the MPI send buffers.
After the MPI communication, the insertion functions {\em copyFromBuf} copy the data 
from the MPI buffers back into the receiving data structures.}
\label{fig::heterogeneous}
\end{figure}
For parallel heterogeneous simulations, the information which Block runs on which hardware
has to be known on all processes in our implementation. Hence, during the initialization we 
set on each process the {\em hs} of all Blocks to the {\em hs} of the process on which they are
allocated.
To determine the {\em hs} of each process, the input for the
simulation describes all possible node configurations and a list which
node belongs to which configuration. A node configuration defines how
many processes can be executed on a particular node and which {\em hs}
should be used for each process.
Using these hardware selectors, it is now possible to utilize
different LBM kernels and simulation data on different compute
architectures.  Further, all compute platforms use an identical layout
for the MPI buffers, which acts as an interface for the MPI
communication. Hence, the data in the MPI buffers is independent of
the underlying hardware. During the MPI parallelization, only the
extraction and insertion function have to be selected according to the
{\em hs} of the Blocks to extract and insert the data from the
different simulation data structures.  Fig.~\ref{fig::heterogeneous}
illustrates this in detail with a heterogeneous LBM simulation.
%

%
%
%
%
%
%
\section{Investigation of Performance and Scalability}
Subsequently, the performance of our design is discussed by
means of Lid Driven Cavity scenarios in $3D$.
In contrast to other highly optimized implementations on \GPUS{} all
measurements presented involve the \PCIe{} data transfer of the
complete halo layer from \CPU{} to \GPU{} and vice versa in each time
step. Therefore, the actual performance is lower in contrast
to~\cite{2008:Toelke:Accuracy} and~\cite{gpu:Habich:2009:paper}.
However, scalability will be rather stable as most of the \PCIe{}
communication time is already accounted for by the single GPU
simulation.
First of all, we investigate the single GPU and CPU performance
including a detailed examination of the overhead for multi-GPU
simulations in Sec.~\ref{sec:multiBlockGPU}. Additionally, the
overhead introduced by several Blocks per process is evaluated to
estimate the suitability of the Patch and Block data structure for
load balancing strategies.
In Sec.~\ref{sec:Multi_perf} we conduct weak and strong scaling
experiments in order to determine how the GPU implementation scales,
on the HPC clusters introduced in Sec.~\ref{sec:hardware}.  Finally, we investigate the performance of
our design for heterogeneous computations in
Sec.~\ref{sec:hetero_perf}.
\subsection{Single GPU and CPU Performance}
\begin{figure}[t!]
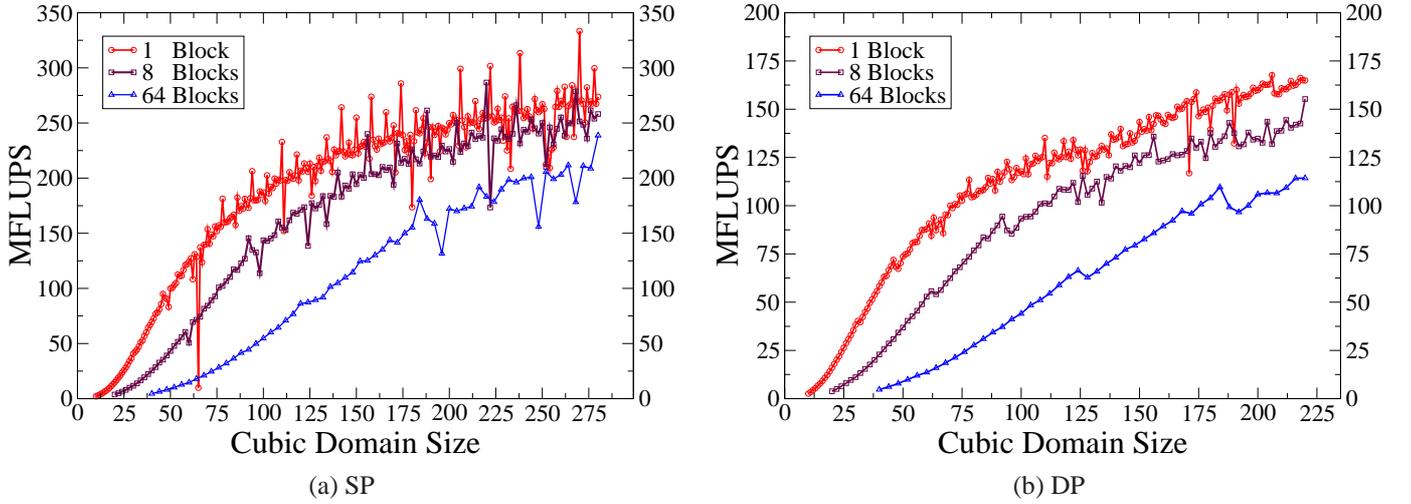

\centerline{\subfloat[SP]{\includegraphics[clip,height=6cm]{images/Results/SingleBlocksSingleGPU-StrongBlockScaling_sp}
\label{plot:MultiBlockSingleGPU_a}
}
\hspace*{2mm}
\subfloat[DP]{\includegraphics[clip,height=6cm]{images/Results/SingleBlocksSingleGPU-StrongBlockScaling_dp}%
 \label{plot:MultiBlockSingleGPU_b}
}}
\caption{Single-\GPU{} measurements on \TINYGPU{} (\TESLA{} C1060) for single (SP) and
            double precision~(DP). A three dimensional partitioning is used to 
            divide the simulation domain into Blocks.
            For a domain size of e.g. $200^3$ lattice cells and $64$ Blocks,
            each Block has a size of $50^3$ lattice cells.}
\label{plot:MultiBlockSingleGPU}
\end{figure} 
\label{sec:multiBlockGPU}
\begin{figure}[b!]
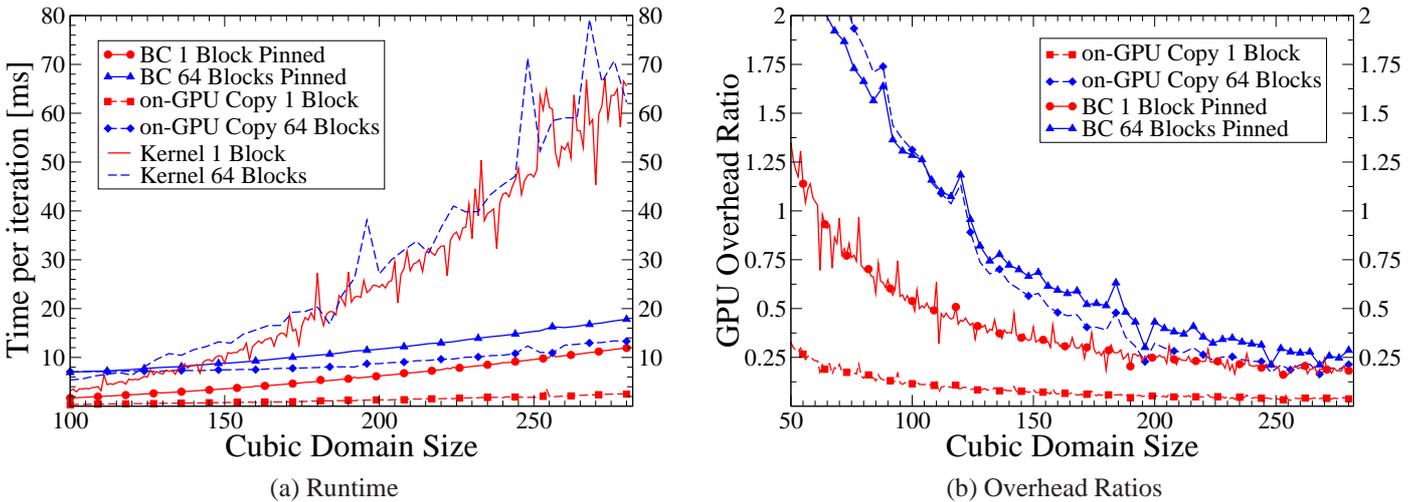

\centerline{
\subfloat[Runtime]{
\includegraphics[clip,height=6cm]{images/Results/SingleBlocksSingleGPU-StrongBlockScaling-Timing_times}
\label{plot:TimeStudyGPU_a}
}
\hspace*{2mm}
\subfloat[Overhead Ratios]{
\includegraphics[clip,height=6cm]{images/Results/SingleBlocksSingleGPU-StrongBlockScaling-Timing_overhead}%
 \label{plot:TimeStudyGPU_b}
}
}
\caption{Single-GPU time measurements on \TINYGPU{}(\TESLA{} C1060) in SP. 
Fig.~(a) shows the runtimes of different parts of the algorithm and Fig.~(b) shows the 
ratio of the times for on-\GPU{} copy operations and boundary condition handling (BC) to the kernel 
execution time. In both Figures results are given for $1$ and $64$ Blocks.
Pinned memory denotes host memory allocated by the \CUDA{} call {\em cudaHostAlloc}.}
\label{plot:TimeStudyGPU}
\end{figure} 
\begin{figure}[htbp]
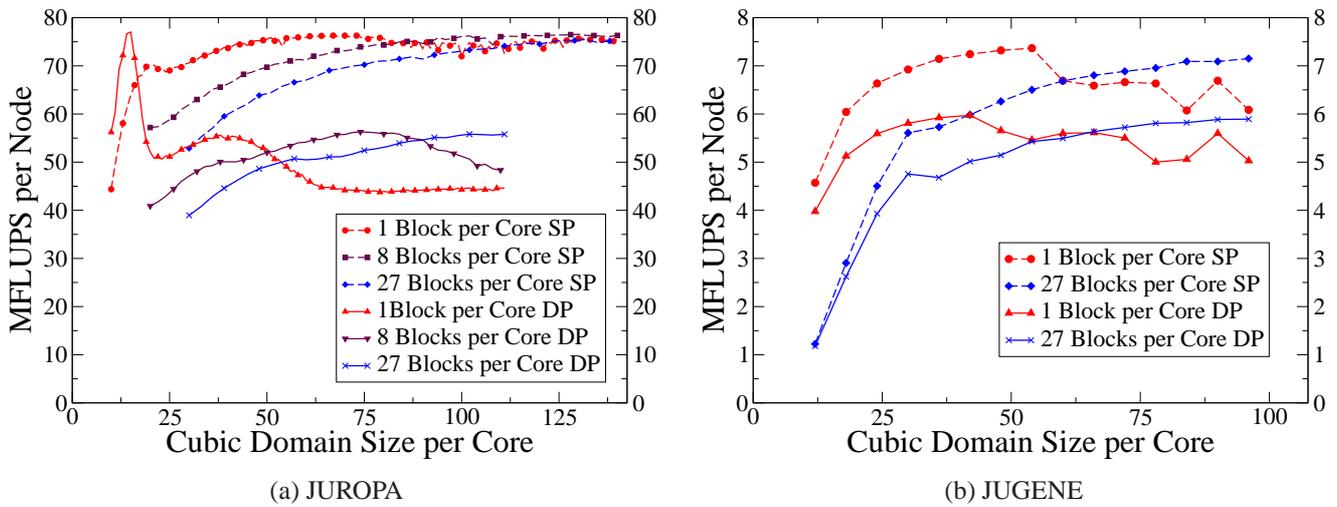

\centerline{
\subfloat[\JUROPA{}]{\includegraphics[clip,height=6cm]{images/Results/SingleBlocksSingleNode-StrongBlockScaling_Juropa}
\label{plot:MultiBlockSingleCPU_a}
}
\hspace*{2mm}
\subfloat[\JUGENE{}]{\includegraphics[clip,height=6cm]{images/Results/SingleBlocksSingleNode-StrongBlockScaling_Jugene}%
 \label{plot:MultiBlockSingleCPU_b}
}
}
\caption{Single node \CPU{} measurements on \JUROPA{} (Xeon X5570) and \JUGENE{} (BlueGene/P) for different Block numbers per core.
For \JUROPA{} $8$ and for \JUGENE{} $4$ cores are used.
}
\label{plot:MultiBlockSingleCPU}
\end{figure} 
Our performance results for a single \GPU{} having $1$ to $64$ local
Blocks are depicted in Fig.~\ref{plot:MultiBlockSingleGPU}. The
performance increases with the domain size and saturates at a domain
size of around $200^3$ lattice cells for a single Block. This is in
contrast to the pure kernel measurements of
Fig.~\ref{plot:MultiBlockSingleGPU-Kernel}, where the maximum performance is already 
reached for a domain size of around $70^3$ lattice
cells. Fig.~\ref{plot:TimeStudyGPU} shows that this results from the
additional overhead of the on-GPU and BC copy operations. The same
holds for the drop in performance using several Blocks, as the pure
kernel runtime of $1$ and $64$ Blocks is nearly identical. For large
domain sizes we loose about $5$~\% for $8$ Blocks and $25$~\% for $64$
Blocks compared to the runtime of one Block. Hence, if several small
Blocks are required, e.g. for load balancing strategies, the performance
of our GPU implementation will be reduced. The maximum
achieved performance is $340$~(SP)~/~$167$~(DP) MFLUPS. Compared to
the pure kernel performance we sustain around $80$~\% using large
domains for both SP and DP. For small domain sizes, e.g. $100^3$
lattice cells, we estimated in Tab.~\ref{tab:HybridComputingEstimate}
a drop in performance from around $300$ to $264$ MFLUPS (SP),
taking only the \PCIe{} transfer into account.  The measurements in
Fig.~\ref{plot:MultiBlockSingleGPU} show a performance of
around $190$ MFLUPS.  As can be seen in Fig.~\ref{plot:TimeStudyGPU},
this discrepancy again results from the, in this case dominating,
overheads of the on-GPU copies and the BC
treatment. This clearly indicates that the \PCIe{} transfer, which is included in the BC treatment, is not
the only component crucial to sustain a large portion of the kernel
performance.  The on-GPU copy operations are hereby unavoidable, but
the BC could be treated directly on the GPUs for further performance
improvement.  This will be investigated in future work.
\par
The single node performance on \JUROPA{} and \JUGENE{} is
presented in Fig.~\ref{plot:MultiBlockSingleCPU}. Compared to the maximum single GPU 
performance the CPU performance corresponds to about $25$~\% in SP and $33$~\% in DP
on \JUROPA{} and $2$~\% in SP and $3.5$~\% in DP on \JUGENE{}.
Usually, we use domain sizes ranging from $90^3$ to $130^3$ in DP on
one CPU core.  For these sizes, the CPU measurements show a superior
performance for multi-Block simulations compared to single
Block simulations.
This is in contrast to the GPU implementation, where
multiple Blocks cause a degradation in performance.  This results from
an efficient utilization of the cache due to blocking effects
occurring especially for the AoS data layout. Hence, for the
investigated architectures block-structured grids are well suited for
load balancing strategies.
\subsection{Multi-\CPU{} and \GPU{} Performance}
\label{sec:Multi_perf}
\begin{figure}[h!]
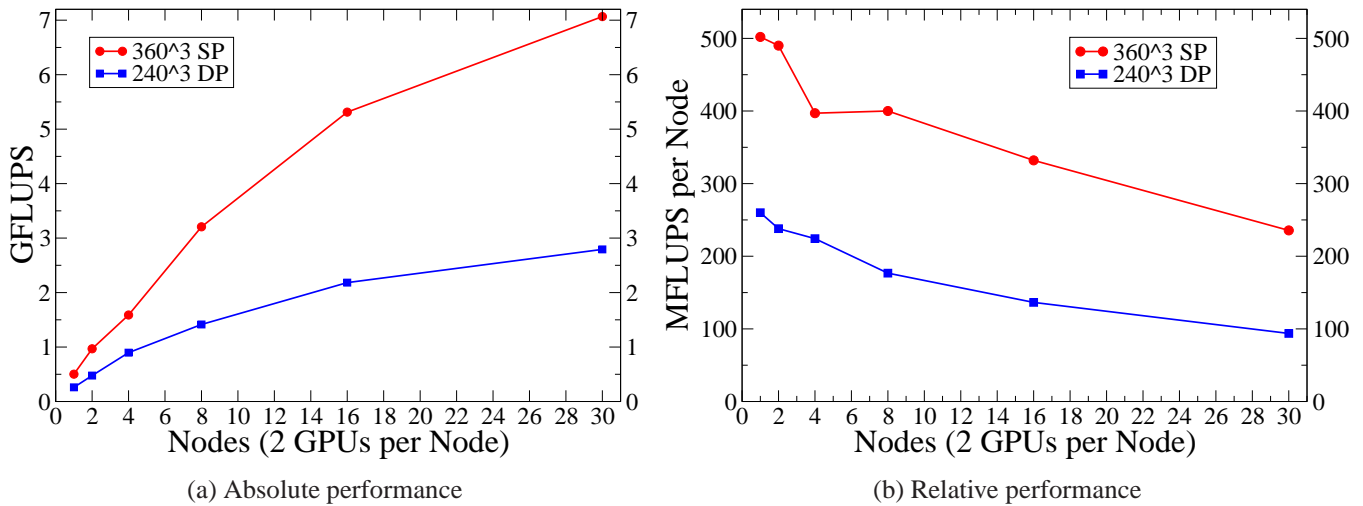

\centerline{
\subfloat[Absolute performance]{
\includegraphics[clip,height=6cm]{images/Results/StrongScalingGPU}
\label{plot:MultiGPUStrongScaling}
}
\subfloat[Relative performance]{
\includegraphics[clip,height=6cm]{images/Results/StrongScalingGPU-perNode}
\label{plot:MultiGPUStrongScalingPerNode}
}
}
\caption{Multi-\GPU{} strong scaling experiments on the \NEC{} (\TESLA{} S1070) with a
  domain size of $360^3$ lattice cells in SP and $240^3$ lattice cells in DP.
  The block decomposition is three
  dimensional. Figure (a) shows the absolute performance values, whereas Figure (b) shows the relative performance, i.e. the absolute performance divided by the number of compute nodes.}
\label{fig:StrongGPU}
\end{figure}  
\begin{figure}[b!]
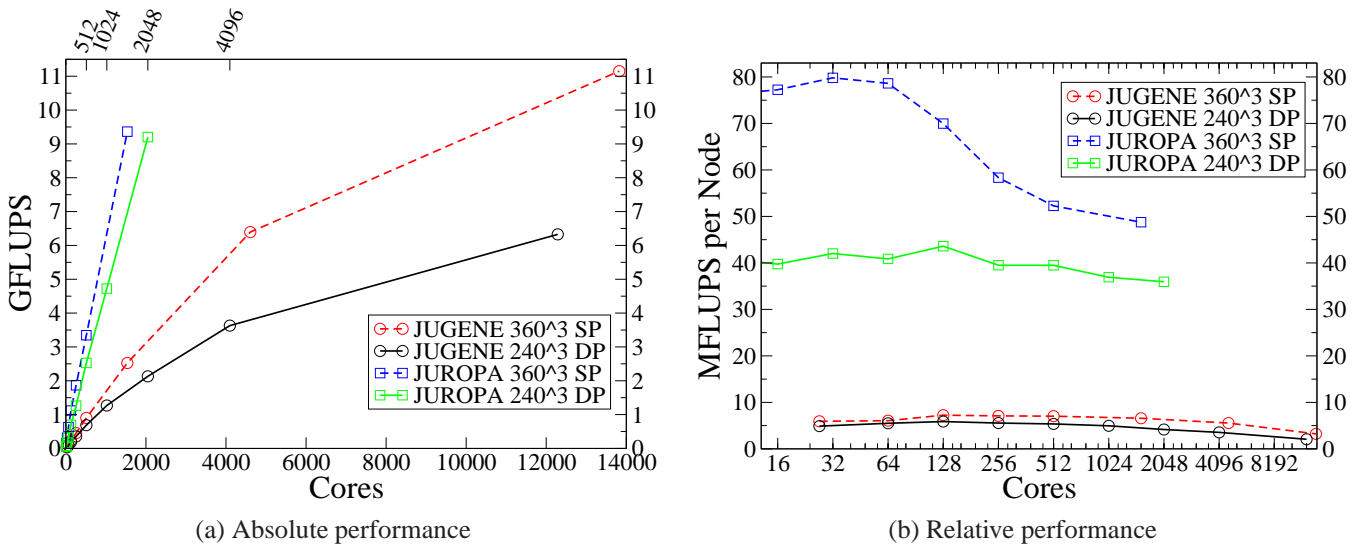

\centerline{
\subfloat[Absolute performance]{\includegraphics[clip,width=0.52\columnwidth]{images/Results/StrongScalingCPU}
\label{plot:MultiCPUStrongScaling}
}
\hspace*{2mm}
\subfloat[Relative performance]{\includegraphics[clip,width=0.52\columnwidth]{images/Results/StrongScalingCPU_perNode}%
 \label{plot:MultiCPUStrongScalingPerNode}
}
}
\caption{Multi-\CPU{} strong scaling performance on \JUROPA{} (Xeon X5570) 
and \JUGENE{} (BlueGene/P). The block decomposition is three dimensional.
}
\label{fig:StrongCPU}
\end{figure}  
There are two basic scenarios to investigate parallel performance:
weak scaling and strong scaling.  In weak scaling experiments, the work
load per compute node is kept constant for an increasing number of
nodes. With this scenario the scalability and the overall manageable
parallelism of the code is evaluated. Strong scaling experiments
answer the question how much the time to solution can be reduced for a
given problem. Therefore, the work load of all nodes is kept constant
leading to a dominating communication overhead and thus a drop in
speedup with an increasing number of nodes.
An important point for the scalability of multi-GPU simulations is
whether the performance scales if using two GPUs on the same node.  On
\TINYGPU{} and the \NEC{} this has been the case, as we achieved
around $95$~\% parallel efficiency for two GPUs. Further, weak scaling
experiments on the \NEC{} showed a nearly linear scaling up to $60$
GPUs for the domain size $222^3$ resulting in a maximum performance of
around $16$ GFLUPS in SP.
In comparison to todays CPUs, single GPUs offer a superior performance.
Hence, on the one hand they should be well suited to
reduce the time to solution in parallel simulations as less internode
parallelism is required. On the other hand, the multi-GPU performance is not
only hampered by the MPI communication, but also by the \PCIe{} transfers, the on-\GPU{} copies, 
and, in contrast to the CPU, the missing cache effect for small domains.
In our GPU strong scaling experiments, depicted in
Fig.~\ref{fig:StrongGPU}, it can be seen that the relative performance 
for $1$ to $30$ compute nodes drops from around $500$ to
$235$ MFLUPS in SP and from around $250$ to $100$ MFLUPS in DP.
Compared to the CPU strong scaling experiments in
Fig.~\ref{fig:StrongCPU}, we need around $6$~(SP~\&~DP) compute nodes
on \JUROPA{} and $75$~(SP)~/~$50$~(DP) on \JUGENE{} to achieve the
performance of a single GPU node on the \NEC{}.  To achieve
the performance of $30$ GPU compute nodes, we need around $137$~(SP)~/~$70$~(DP)
compute nodes on \JUROPA{} and $1275$~(SP)~/~$750$~(DP) on
\JUGENE{}. The corresponding parallel efficiencies are: $46$~(SP)~/~$37$~(DP)~\%
for the GPU implementation on \NEC{}, $65$~(SP) /
$93$~(DP)~\% for the CPU implementation on \JUROPA{} and $90$~(SP)~/~$98$~(DP)~\%
on \JUGENE{}. Hence, to achieve the same time to
solution our GPU implementation makes less efficient use of the
utilized hardware, but also requires fewer nodes.
\subsection{Heterogeneous \GPU{}--\CPU{} Performance}
\label{sec:hetero_perf}

\begin{figure}[htbp]
\centerline{
\subcaptionbox{
Investigation of the load balance with 6 CPU-only processes each working on one Block
and $2$ GPU processes with varying block counts. The Block size is $90^3$ lattice cells.
\label{plot:HeteroLaodBalance}}{
\includegraphics[clip,height=5.5cm]{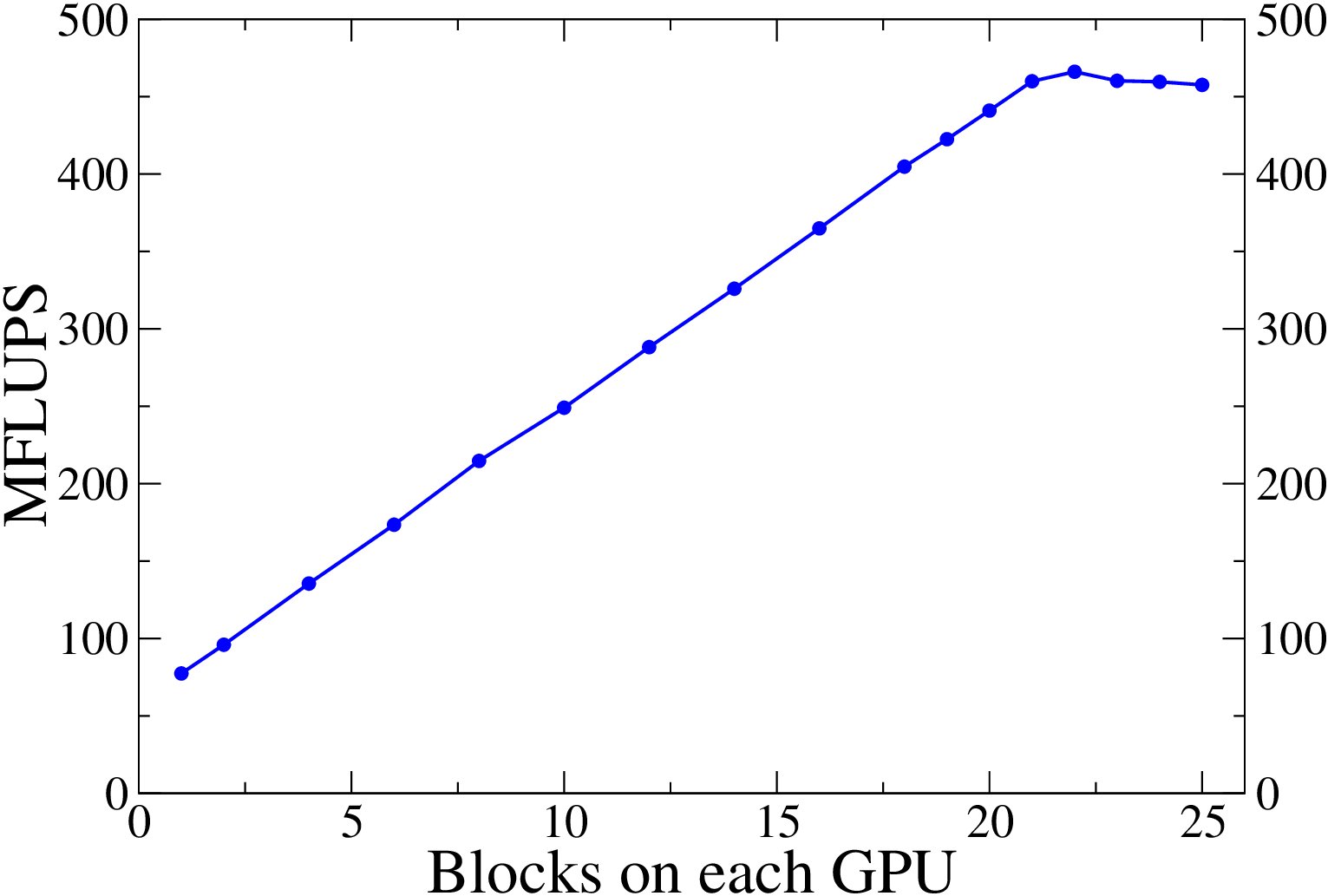}
}
\hspace*{2mm}
\subcaptionbox{Performance comparison between homogeneous and heterogeneous setups depending on the Block size.\label{tab:hetero}}{
\footnotesize
\begin{tabular}[b]{lllll}\toprule[0.3mm]
 Block Size& $70^3$ & $71^3$ & $90^3$& $91^3$ \\
 Blocks & 44 & 44 & 50 &50\\
\midrule
Processes&&&&\\
2 x GPU &379.1&341.6&422.6&404.3\\
2 x GPU + 6 x CPU&423.2&382.6&466.7&446.1\\
\midrule[0.3mm]\\
\midrule[0.3mm]
 Block Size& $70^3$ & $71^3$ & $90^3$& $91^3$ \\
\midrule
Processes&&&&\\
6 x CPU (6 Blocks)& 58.5&58.6 &58.1 & 58.1 \\
2 x GPU (2 Blocks)&388.2&431.2&495.3&469.2 \\
\bottomrule[0.3mm]
\\[-1mm]
\end{tabular}
}
}
\caption{Heterogeneous performance on one compute node of \TINYGPU{} (\TESLA{} C1060) in SP.}
\end{figure}
To discuss the capabilities of heterogeneous simulations on GPUs and
CPUs we first investigate the performance on a single compute node of
\TINYGPU{}.  Here, best performance results are achieved with $6$ CPU
only processes and $2$ for the GPUs. Additionally, the work load for
each process has to be adjusted. This is depicted in
Fig.~\ref{plot:HeteroLaodBalance}, where each CPU process has one
Block with $90^3$ lattice cells, whereas the number of blocks
allocated on each GPU process is increased until the work load is
balanced. Note that in the load-balanced case of $22$ Blocks on each GPU,
the runtime of the GPU kernel is still $33$~\% lower than the runtime of the
CPU kernel, as on the GPU side a larger communication overhead is added to overall
runtime.
Further, in Tab.~\ref{tab:hetero} the node performance of
heterogeneous simulations is compared to simulations using only GPUs
having the same number of Blocks or just one Block on each
GPU. Hereby, the number of Blocks is chosen so that the
heterogeneous simulations are load balanced.  It can be seen that the
heterogeneous simulations yield an increase in performance of
around $42$ MFLUPS for all Block sizes, whereas the maximum for $6$
CPU processes would be around $58$ MFLUPS. Compared to simulations
running on two GPU processes, which have only one Block on each
process we loose around $5 - 12 $~\% performance due to the
increased overhead. For the $70^3$ Block size the kernel 
performance is overly high due to padding effects and hence we gain
around $10$~\% in performance.
Summarizing, for the mere purpose of a performance increase
our current heterogeneous implementation is not suitable, but for
simulations requiring several blocks on each process, e.g. for load
balancing strategies or other optimizations, it is possible to improve
the performance. Additionally, with our implementation the
memory of GPU and CPU can be utilized, which allows for larger
simulation setups.
So far, we have only considered heterogeneous simulations on a single compute
node. In Tab.~\ref{tab:hetero:scaling} weak scaling experiments up to
$90$ compute nodes are depicted. The weak scaling experiment using
$60$ GPUs on $30$ compute nodes shows a perfect parallel efficiency
and the heterogeneous experiment running on $60$ GPUs and $180$ CPU
only processes has a parallel efficiency of $96$~\%. In addition, we
have conducted scaling experiments using different kind of compute
nodes, e.g. compute nodes having only a CPU and nodes having
additional GPUs. It can be seen that the performance scales
well from $30$ up to $90$ compute nodes.  Hence, with our
implementation it is possible to efficiently utilize all nodes on
clusters having heterogeneous node configurations. A further
improvement of our heterogeneous design for multiphysics simulations
could be the simulation of complex spatially contained functionality,
e.g. a rising bubble, on processes running on CPUs and to only simulate
pure fluid regions on the GPUs, for which they are currently
suited best.
\begin{table}
\footnotesize
\centering
\begin{tabular}{lcclllll}\toprule[0.3mm]
Blocks&\multicolumn{2}{c}{GPU: $1$}&& \multicolumn{4}{c}{GPU: $22$, CPU: $1$}\\
\cmidrule{1-3}
\cmidrule{5-8}
Nodes &1&30&&1&30&60&90\\
Processes & 2 x GPU & 60 x GPU&& 2 x GPU + & 60 x GPU + &  60 GPU + & 60 GPU + \\
&  & & & 6 x CPU & 180 x CPU &  420 x CPU & 660 x CPU\\
\cmidrule[0.3mm]{1-3}
\cmidrule[0.3mm]{5-8}
MFLUPS & 476 & 14480 & & 459 & 13267 & 15684 & 17846 \\
\bottomrule[0.3mm]
\end{tabular}\\[4mm]
\caption{Heterogeneous weak scaling experiments using up to $90$ compute nodes on the \NEC{}. The simulation 
domain for nodes with GPUs is $90x4500x90$ and for CPU-only nodes $90x540x90$. All
presented results are in SP and the load 
for the heterogeneous simulations is balanced.}
\label{tab:hetero:scaling}
\end{table}

%
%
%
%
%
%

\section{Conclusion}
\label{sec:Conculsion}
%
%
A fundamental requirement for the utilization of GPUs in HPC clusters
are scalable multi-GPU implementations.  In this article, we have
shown that this is possible for the LBM.  Additionally, by means of
our Patch and Block design, and our functionality management we have
presented an approach for heterogeneous simulation on clusters
equipped with varying node configurations.  Further, we have shown
that with our WaLBerla framework good runtime performance results can
be achieved on various compute platforms despite the overhead for
flexibility and its suitability for multiphysics simulations.  In
future work, we will optimize the memory accesses of our GPU
implementation with the help of padding strategies as well as
implement arbitrary boundary conditions, directly computed on the GPU.
We will also investigate hybrid OpenMP and MPI parallelization in
combination with heterogeneous simulations.

\section*{Acknowledgments}
This work is partially funded by the European Commission with \emph{DECODE}, CORDIS 
project no. 213295, by the Bundes\-ministerium f{\"u}r Bildung und Forschung under
the \emph{SKALB} project, no. 01IH08003A, as well as by the ``Kompetenznetzwerk
f{\"u}r Tech\-nisch\--Wis\-sen\-schaft\-liches Hoch- und H{\"o}chst\-leistungs\-rechnen
in Bayern'' (\emph{KONWIHR}) via \emph{waLBerlaMC}. Compute resources on \JUGENE{} and \JUROPA{} 
were provided by the 
John-von-Neumann Institute (Research Centre J\"ulich) under the HER12 project. 
We thank the DEISA Consortium,
co-funded through the EU FP6 project RI-031513 and the FP7 project
RI-222919, for support and access to Juropa within the DEISA
Extreme Computing Initiative. Access to the systems at HLRS was granted through
\emph{Bundesprojekt LBA-Diff}. 

\bibliographystyle{elsarticle-num}
\bibliography{IEEEabrv,bibfile}

\end{document}